# Vortex states in submicron titanomagnetite particles

O. N. Serebryakova, N. A. Usov

*Pushkov Institute of Terrestrial Magnetism, Ionosphere and Radio Wave Propagation, Russian Academy of Sciences, IZMIRAN, 108840, Troitsk, Moscow, Russia*

**Abstract.** Numerical simulation is used to study the magnetic properties of spheroidal titanomagnetite particles $Fe_{3-x}Ti_xO_4$ ($0.4 \leq x \leq 0.6$) with semi-axes ratio ranging from 0.5 to 1.5 and transverse diameters $D$ = 100 - 480 nm. It is shown that various vortex micromagnetic states can emerge in the titanomagnetite particles depending on the titanium concentration, aspect ratio and particle diameter. The single-domain particle diameters are determined using the energy diagrams for stationary uniform and vortex states. The remanent magnetization of dilute assemblies of titanomagnetite particles with varying titanium content are calculated as a function of the transverse particle diameter $D$. It is demonstrated that for submicron-sized titanomagnetite particles, $D \sim 500$ nm, the normalized remanent magnetization of the assembly reaches very low values, $M/M_s < 0.01$, where $M_s$ is the saturation magnetization of the particles.

**Keywords:** titanomagnetite, vortex state, single-domain diameter, remanent magnetization, numerical simulation

**Corresponding author:** O. N. Serebryakova

E-mail: silver@izmiran.ru

## 1. Introduction

Assemblies of iron oxide magnetic particles with diameters comparable to or greater than the single-domain threshold have recently attracted attention due to their applications in biomedicine [1–3]. The study of the properties of non-uniform micromagnetic states in submicron-sized magnetite and titanomagnetite particles is also essential for the accurate interpretation of extensive paleomagnetic and geophysical datasets. It is well established [4–6] that ancient volcanic rocks often contain dispersed assemblies of titanomagnetite particles, whose remanent magnetization preserves unique information regarding the magnitude and direction of the Earth's global magnetic field in the distant past. This information is of great scientific value, enabling, for instance, the dating of tectonic plate movements, the determination of time intervals for geomagnetic polarity reversals, the refinement of the age of archaeological sites, etc. Notably, in natural titanomagnetite particle assemblies, the proportion of single-domain (SD) particles with diameters $D \leq 200$ - 250 nm which are capable of maintaining a large stable remanent magnetization at room temperature is typically small. Consequently, a question arises concerning the origin of the long-term remanent magnetization observed in ancient volcanic rocks that contain a significant number of submicron-sized particles. It has been hypothesized [7, 8] that the exchange cores of single-vortex (SV) micromagnetic states within submicron iron oxide particles may make a substantial contribution to the remanent magnetization of volcanic rocks.

Micromagnetic calculations of non-uniform magnetization distributions in particles of pure magnetite, $Fe_3O_4$, have been performed over a fairly wide range of particle diameters [6, 9–17]. It has been shown both theoretically and experimentally [6, 9–16] that submicron magnetite particles exhibit SV states, with their remanent magnetization decreasing sharply as the particle diameter increases. In spherical magnetite particles of sufficiently large diameter these vortex states transform into two types of circular domain walls with similar total energies, having rotation axis aligned parallel to either an easy axis or a hard axis of the cubic magnetic anisotropy of the magnetite, respectively [17]. Similar structures also exist in spheroidal magnetite particles of sufficiently large sizes.

The micromagnetic structure of submicron titanomagnetite particles has been studied in less detail [18,19], as the magnetic parameters of these particles depend significantly on their titanium concentration. The chemical composition of titanomagnetite particles dispersed in volcanic rocks is described by the formula $Fe_{3-x}Ti_xO_4$, $(0 \leq x \leq 1)$, where $x$ represents the titanium fraction [4]. Following standard notation [19], particles with a titanium content of $x = 0.4$ will henceforth be referred to as TM40. Similar notation is used for other compositions.

Submicron titanomagnetite particles with compositions ranging from TM40 to TM60 constitute a significant fraction of the particles dispersed within volcanic rocks [4–8]. As the titanium content increases within the range of $x = 0.4$–$0.6$, the particle saturation magnetization $M_s$ decreases substantially. In addition, the cubic magnetic anisotropy constants $K_{1c}$ and $K_{2c}$, and even the orientation of the magnetic easy anisotropy axes undergo noticeable changes [19]. Although the shape of titanomagnetite particles in natural assemblies is often quite irregular, investigating the magnetic properties of spheroidal particles [19] serves as a necessary step toward understanding how the magnetic properties of titanomagnetite particles evolve in response to changes in their composition, size, and shape.

When analyzing the magnetic properties of titanomagnetite particle assemblies it seems appropriate to distinguish three characteristic particle size ranges. These include: SD particles with sizes below 200–250 nm; particles ranging from 200 to 500 nm, in which SV states possess the lowest total energy; and, finally, particles on the order of 1 μm or larger, which exhibit multi-vortex or multi-domain configurations [18–20]. The magnetic properties of SD particles, which could potentially be the primary contributors to the remanent magnetization of an assembly, have been studied most extensively to date [21–24]. However, their proportion in ancient igneous rocks is relatively small. At the same time, SV states can exhibit significant remanent magnetization [18, 19], substantially exceeding that of multi-vortex or multi-domain configurations.

In addition to the results of Refs. 18, 19, in this work energy diagrams of stationary SV states realized in spheroidal titanomagnetite particles of composition TM40 – TM60 are constructed. Calculations are carried out by solving numerically the dynamic Landau-Lifshitz-Gilbert (LLG) equation [25-28] for spheroidal particles with the semi-axes ratio $a/b = 0.5 – 1.5$ in the range of transverse particle diameters $D = 100 – 480$ nm. It is shown that, depending on the titanium concentration and aspect ratio, various types of SV states are realized in titanomagnetite particles, which compete in energy with the uniform magnetization. The single-domain diameters of titanomagnetite particles were assessed depending on the aspect ratio and titanium content. It is shown that in the indicated range of diameters, the reduced remanent magnetization of dilute assemblies of titanomagnetite particles decreases in an inverse power-law manner as a function of the particle diameter and takes on very small values, $M/M_s < 0.01$, at $D \sim 500$ nm.

## 2. Numerical simulation

In this work stable micromagnetic states in submicron titanomagnetite particles are determined by solving the dynamic LLG equation for the particle's unit magnetization vector $\alpha(\vec{r})$ [25–28]

$$\frac{\partial \vec{\alpha}}{\partial t} = -\gamma\left(\vec{\alpha} \times \vec{H}_{ef}\right) + \kappa\left(\vec{\alpha} \times \frac{\partial \vec{\alpha}}{\partial t}\right), \tag{1}$$

where $\gamma$ is the gyromagnetic ratio, and $\kappa$ is the magnetic damping constant. The effective magnetic field $\vec{H}_{ef}$ acting on a unit magnetization vector is calculated as the derivative of the total particle energy $W$, which is the sum of the exchange energy $W_{ex}$, the magnetic anisotropy energy $W_a$ and the magnetostatic energy $W_m$ [21,22]

$$\vec{H}_{ef} = -\frac{\partial W}{V M_s \partial \vec{\alpha}}, \qquad M_s \vec{H}_{ef} = C\Delta\vec{\alpha} - \frac{\partial w_a}{\partial \vec{\alpha}} + M_s \vec{H}'. \tag{2}$$

Here $V = 4\pi ab^2/3$ is the volume of the spheroidal particle, $M_s$ is the saturation magnetization, $C = 2A$ is the exchange constant, $w_a(\vec{\alpha}) = W_a/V$ is the energy density of magneto- crystalline anisotropy, and $\vec{H}'$ is the vector of the demagnetizing field acting in the volume of the particle. The demagnetizing field is calculated based on the exact expression for the interaction energy of uniformly magnetized cubes of finite size [22, 25].

The spheroidal titanomagnetite particles studied in this work are assumed to be either prolate or oblate along the Z-axis of a Cartesian coordinate system, where $a$ and $b$ are the longitudinal and transverse semi-axes of the spheroid, respectively, and $D = 2b$ is the transverse particle diameter. Thus, depending on the aspect ratio $a/b = 0.5$, 0.75, 1.0, 1.25 and 1.5, the $Z$-axis serves as either the easy or hard axis for the shape anisotropy energy of a single-domain particle.

Although, in the general case, the orientation of the easy axes of cubic magneto-crystalline anisotropy may be arbitrary with respect to the particle's symmetry axis, this study employs the standard expression for the cubic magnetic anisotropy energy density of the particle [19, 29]

$$w_a = K_{1c}\left(\alpha_x^2\alpha_y^2 + \alpha_x^2\alpha_z^2 + \alpha_y^2\alpha_z^2\right) + K_{2c}\alpha_x^2\alpha_y^2\alpha_z^2, \tag{3}$$

where $K_{1c}$ and $K_{2c}$ are the cubic magneto-crystalline anisotropy constants of titanomagnetite. The values of magnetic constants used in the calculations for TM40–TM60 particles are taken from [19] and are presented in Table 1.

Table 1. Magnetic parameters of titanomagnetite particles [19]

| | $M_s$ (emu/cm$^3$) | $K_{1c}$ (erg/cm$^3$) | $K_{2c}$ (erg/cm$^3$) | $A$ (erg/cm) | $L_{ex} = \sqrt{2A}/M_s$ (nm) |
|---|---|---|---|---|---|
| TM40 | 240.0 | $-1.3\times10^5$ | $0.18\times10^5$ | $0.93\times10^{-6}$ | 56.8 |
| TM50 | 190.0 | $-5.8\times10^4$ | $0.9\times10^5$ | $0.8\times10^{-6}$ | 66.5 |
| TM60 | 140.0 | $7.7\times10^3$ | $2.3\times10^5$ | $0.65\times10^{-6}$ | 81.4 |

Thus, as the titanium content increases within the range $x = 0.4–0.6$, the saturation magnetization of the particles $M_s$ and the exchange constant $A$ decrease; the first cubic anisotropy constant $K_{1c}$ also decreases in absolute value, changing sign for the TM60 composition, while the second constant $K_{2c}$ increases significantly. Since the constant $K_{1c}$ is negative for TM40 and TM50 particles, their easy axes of cubic anisotropy are oriented parallel to the vectors (±1/√3, ±1/√3, ±1/√3). Conversely, for particles with the TM60 composition, which have a positive $K_{1c}$, these axes are parallel to the Cartesian coordinate axes.

As shown in Table 1, the exchange length $L_{ex}$, i.e. the characteristic scale over which the unit magnetization vector can rotate within the particle volume, ranges from 56.8 to 81.4 nm for particles with compositions TM40–TM60, respectively. For numerical modeling a spheroidal particle is approximated by an assembly of small ferromagnetic cubes with a side length significantly smaller than the exchange length of the ferromagnetic material. In the calculations performed the sizes of the cubic numerical cells were generally between 7.5 and 15 nm, reaching 25 nm only for particles with the largest transverse diameters. To approximate non-uniform magnetization distributions within a particle with sufficient accuracy, approximately $N \sim 10^4$ numerical cells are required. The magnetic damping parameter is set to $\kappa = 0.5$.

In accordance with Eq. (1), it is assumed that the final magnetization distribution within a particle is stable, provided that

$$\max_{(1\le i\le N)} \left|\vec{\alpha}_i \times \vec{H}_{ef,i}\right| \Big/ \left|\vec{H}_{ef,i}\right| < 10^{-6}, \qquad (4)$$

where $\vec{\alpha}_i$ and $\vec{H}_{ef,i}$ are the unit magnetization vector and the effective magnetic field vector in the $i$-th numerical cell, respectively.

To identify all stable micromagnetic states that exist within a given particle for a specific range of sizes, it is necessary to perform a sufficient number of calculations, starting from various initial magnetization distributions. In particular, it is useful to employ random initial states, where the orientations of the unit magnetization vectors in the individual numerical cells, $i = 1 - N$, are assigned completely arbitrarily at the initial time. It follows from Eq. (1) that the total particle energy reduces during temporal evolution of initial magnetization distribution, [21]. Furthermore, the total energy of any stationary micromagnetic state is bounded from below. Consequently, by iterating Eqs. (1) and (2) with a sufficiently small time step, a stationary magnetization distribution, namely the one that satisfies criterion (4), is reached within a finite number of steps. It should be noted that the set of stationary states, obtained for a given particle by solving the dynamic LLG equation, is independent of the chosen value for the magnetic damping constant $\kappa$, as the criterion for a stationary micromagnetic state, Eq. (4) does not depend on $\kappa$.

## 3. Energy diagrams of stationary micromagnetic states

The calculations performed for titanomagnetite particles with transverse diameters in the range $D$ = 100 – 480 nm have revealed a large number of stationary micromagnetic states, including both uniform and vortex-type configurations. When analyzing these results one has to take into account that by virtue of Eq. (3) the total energy of a spheroidal titanomagnetite particle is invariant under a 90° rotation about the particle's symmetry axis $Z$ and under a sign reversal of the longitudinal component of the unit magnetization vector, $\alpha_z \rightarrow -\alpha_z$.

Let us first consider the uniform magnetization states. For oblate particles with aspect ratio $a/b$ = 0.5, the shape anisotropy energy determines the direction of the uniformly magnetized state. Consequently, for SD particles of compositions TM40 and TM50, given that $K_{1c} < 0$, the particle magnetization lies within the $XY$ plane, oriented along the equivalent directions (±1/√2, ±1/√2, 0). The states of this type will hereafter be denoted as uxy. The magnetization of TM60 particles with $K_{1c} > 0$ is oriented along the equivalent directions (±1, 0, 0) and (0, ±1, 0); such states will hereafter be denoted as ux.

SD particles with aspect ratio $a/b$ = 0.75 and compositions TM50 and TM60 exist in uxy and ux states, respectively. Particles with composition TM40 are magnetized in directions equivalent to the vector $\left(\xi, \xi, \pm\sqrt{1-2\xi^2}\right)$ where $\xi = \pm 0.66$. Homogeneous states of this type will henceforth be denoted as ud.

Spherical titanomagnetite particles, $a/b$ = 1, in a homogeneous state are magnetized along the corresponding easy axes of magneto-crystalline anisotropy, Eq. (3). For compositions TM40 and TM50 these are ud states with $|\xi| = 1/\sqrt{3}$, while for TM60 composition, they are ux states, as well as states magnetized along the $Z$ -axis, which will be denoted as uz.

SD particles with $a/b$ = 1.25 of compositions TM40 and TM50 are in ud states with values $|\xi|$ = 0.516 and 0.444, respectively. For the TM60 composition such particles are in uz states. Finally, elongated TM40 particles with $a/b$ = 1.5 are in ud states with $|\xi|$ = 0.452, whereas elongated TM60 particles are in uz states. Particles of the TM50 composition are magnetized in directions equivalent to the vector (0.478, 0, 0.878). These states will be denoted as uxz.

It is convenient to classify SV states in spheroidal titanomagnetite particles in a similar way, taking into account the direction of the reduced magnetic moment of the particle, $\boldsymbol{m} = \langle \boldsymbol{M} \rangle / M_s$, which is close to the direction of the approximately uniformly magnetized core of the vortex. Fig. 1 shows the general view of magnetization distributions for SV states in spheroidal titanomagnetite particles of compositions TM40 – TM60 with different aspect ratios in the range $a/b$ = 0.5 – 1.5.

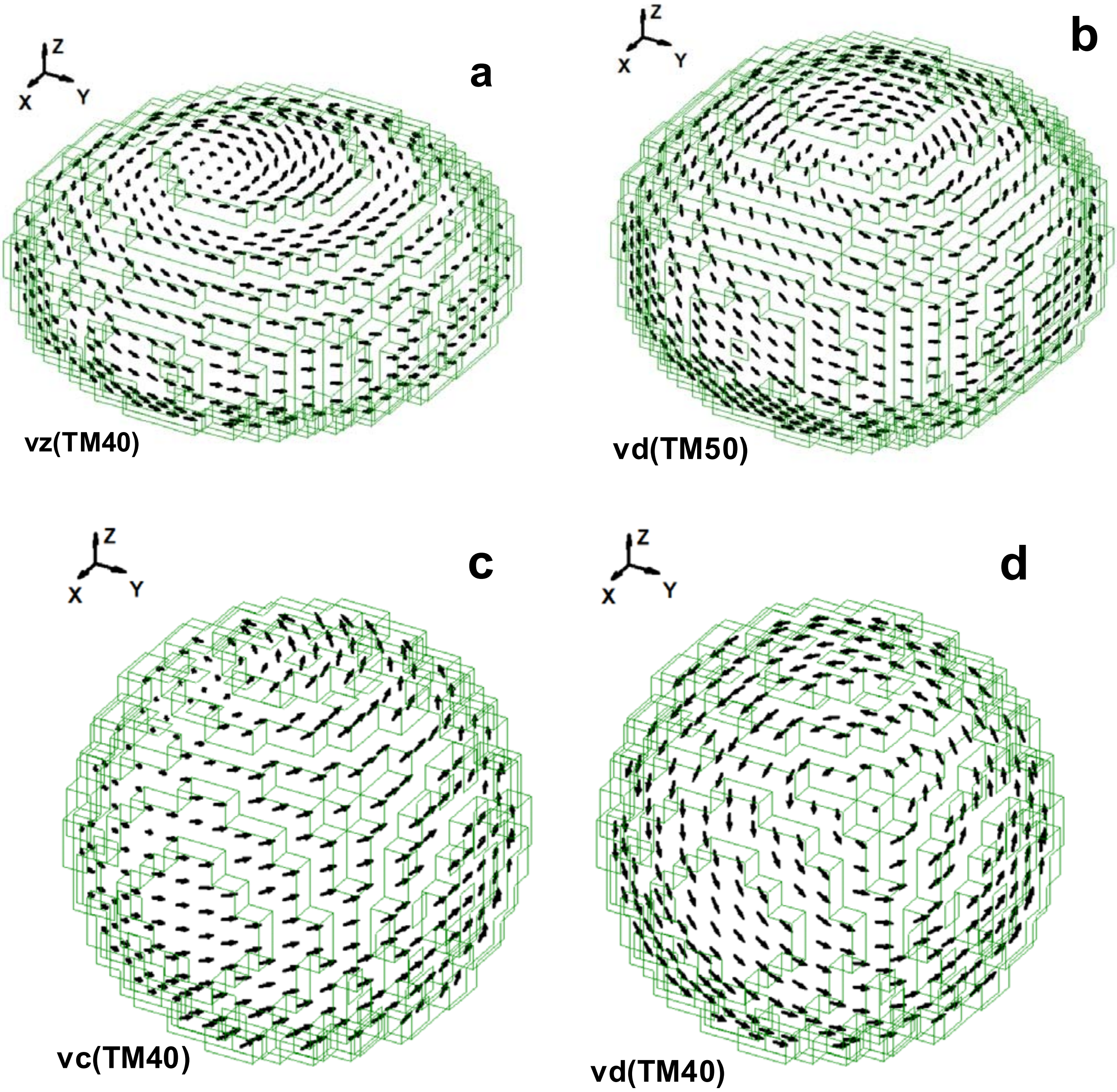

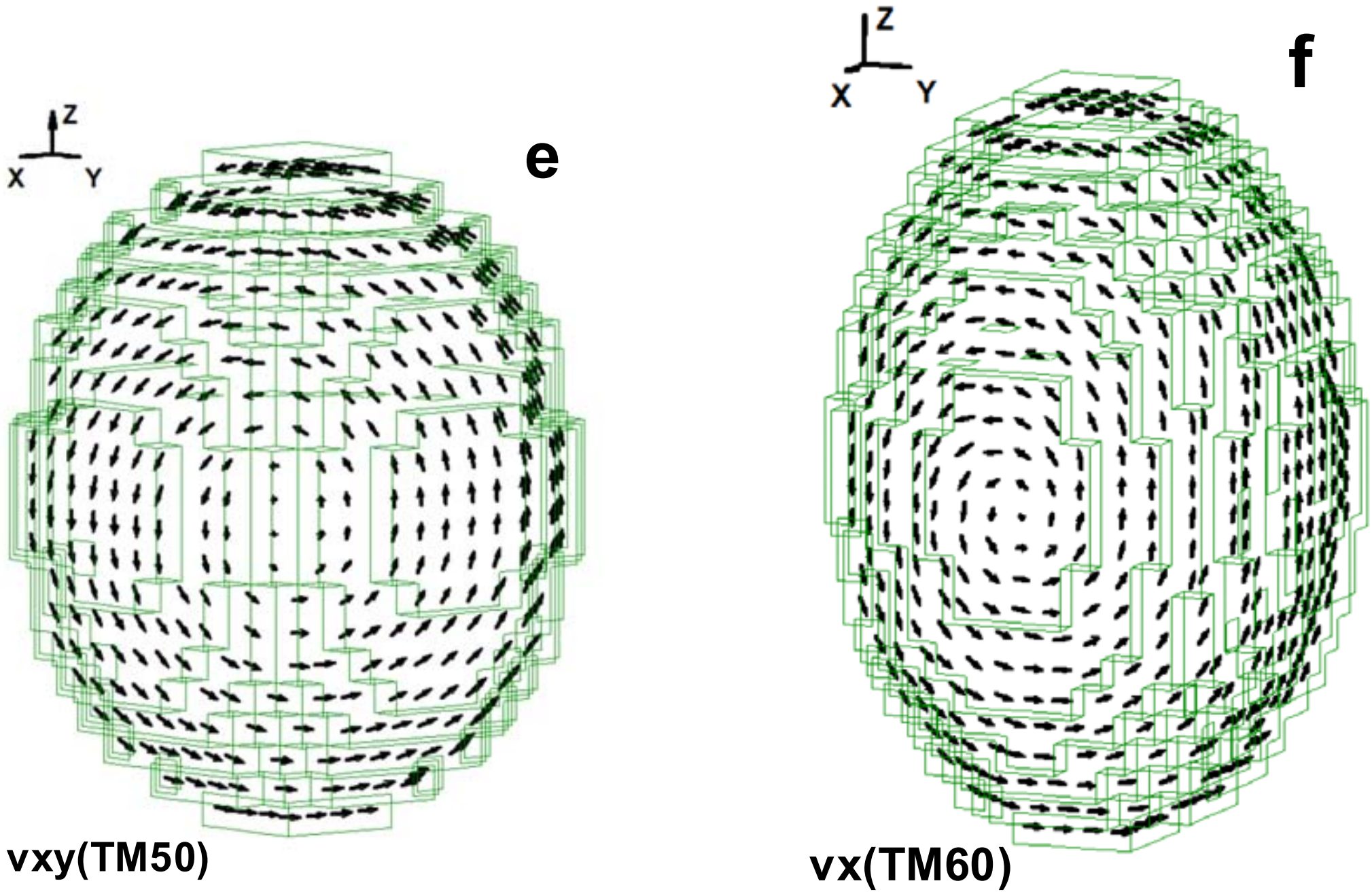

Fig. 1. General view of magnetization distributions for vortex states in titanomagnetite spheroidal particles of various compositions and aspect ratios: a) $a/b$ = 0.5, $D$ = 320 nm; b) $a/b$ = 0.75, $D$ = 400 nm; c) $a/b$ = 1.0, $D$ = 200 nm; d) $a/b$ = 1.0, $D$ = 280 nm; e) $a/b$ = 1.25, $D$ = 220 nm; f) $a/b$ = 1.5, $D$ = 480 nm.

For oblate particles with aspect ratio $a/b$ = 0.5, beyond the single-domain limit only the vz state occurs for all compositions from TM40 to TM60. The core of this vortex is magnetized strictly along the $Z$ -axis, while the outer part rotates around this axis within the $XY$ -plane. The general view of the vz state, with a reduced magnetic moment of $\boldsymbol{m}$ = (0, 0, 0.0575), is shown in Fig. 1a for a TM40 particle. In the case of $a/b$ = 0.75, for TM40 and TM50 compositions vd vortices also emerge in addition to vz vortices as the particle diameter increases. The general view of the vd state, with a reduced magnetic moment $\boldsymbol{m}$ = (0.0443, 0.0443, 0.0739) is shown in Fig. 1b. An important feature of vd states is that the magnitudes of the magnetic moment projections onto the $X$ and $Y$ axes are equal $|m_x| = |m_y|$, and the longitudinal projection is non-zero, $|m_z| \neq 0$.

For spherical particles, $a/b$ = 1.0, vc and vd vortices are realized beyond the single-domain boundary for all compositions TM40 – TM60. Unlike the vz state, the magnetic moment of the vc vortices can be directed along any of the Cartesian coordinate axes, all these states being equivalent. The magnetic moments of vortices vd in the case of spherical particles are parallel to the directions ($\pm1/\sqrt{3}$, $\pm1/\sqrt{3}$, $\pm1/\sqrt{3}$). Thus, they have equal modules of projections on the X, Y and Z axes. Fig. 1c shows the vc state with reduced moment $\boldsymbol{m}$ = (0, 0, 0.476), and Fig. 1d shows the vd state with moment $\boldsymbol{m}$ = (0.075, 0.075, 0.075).

Fig. 1e shows the general view of the vxy vortex with a reduced magnetic moment $\boldsymbol{m}$ = (0.184, 0.184, 0.0) in a particle with aspect ratio $a/b$ = 1.25. The vxy state is similar to the vd state, with the difference that the projection of the particle's magnetic moment onto the symmetry axis $Z$ is strictly zero, while the magnitudes of the projections onto the $X$ and $Y$ axes remain equal. Finally, Fig. 1f shows the general view of the vx vortex in TM60 particle with aspect ratio of $a/b$ = 1.5 with a reduced magnetic moment $\boldsymbol{m}$ = (0.067, 0, 0). The magnetic moment of vx vortices is parallel to either $X$ or $Y$ Cartesian axis, as these directions serve as the easy anisotropy axes for TM60 particles.

The micromagnetic structures of vz and vd vortices are described in detail in [17], using spherical magnetite particles as an example. The micromagnetic structure of the vx state, which occurs in elongated TM60 particles, is illustrated in Fig. 2, which shows two central cross-sections of the particle. As Fig. 2 demonstrates, in the vx state, the unit magnetization vector rotates around the X axis (or the equivalent Y axis), and the magnetic field lines in the outer region of the vortex exhibit significant elliptical deformation.

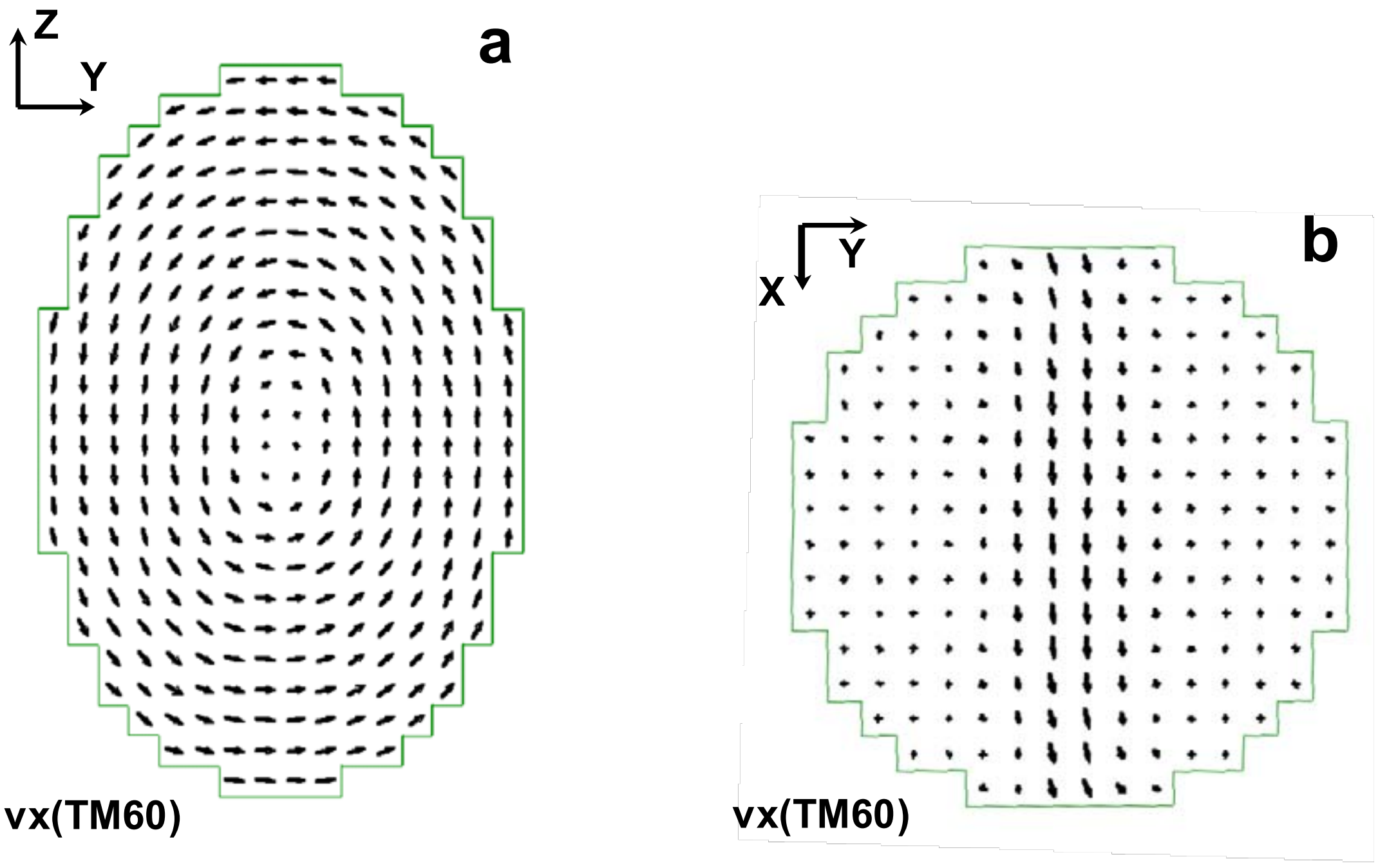

Fig. 2. Vortex magnetization distribution for the vx state in a titanomagnetite particle of composition TM60, with an aspect ratio $a/b$ = 1.5 and a transverse diameter $D$ = 400 nm: (a) central cross-section of the particle in the $YZ$ plane; (b) the same in the $XY$ plane.

Fig. 3 presents energy diagrams for stationary micromagnetic states that show the dependence of the total energy density of spheroidal titanomagnetite particles in various stable micromagnetic states on their transverse diameters. These diagrams make it possible to determine the existence ranges for different micromagnetic states and to estimate the transverse single-domain diameters of the particles. These values are found at the intersection point of the respective curves for the uniform and vortex states [21]. It should be noted that some of the stationary states we discovered, which exist only within narrow ranges of particle diameters, are not shown in the diagrams in Fig. 3. Fig. 3 illustrates a well-known fact [21, 30]: the total energy density of a uniform state is independent of the size of a spheroidal particle, whereas the total energy density of a vortex state decreases rapidly as the particle diameter increases.

The most simple is the energy diagram for oblate particles, $a/b$ = 0.5, shown in Fig. 3a. In this case the single-domain diameter is determined by an energy competition between the vz state, and various uniform states. Specifically, these are the uxy states for TM40 and TM50 compositions, and the ux state for TM60 composition. As illustrated in Fig. 3b, the same situation applies to TM60 particles with aspect ratio $a/b$ = 0.75. On the other hand, for TM40 and TM50 particles two distinct vortex states, namely vz and vd, can occur. In the case of the TM40 composition the vz state competes in energy with the uniform state ud near the single-domain diameter, $D_c$ = 160 nm. However, vz state subsequently becomes unstable and disappears within the diameter range 200 ≤ $D$ ≤ 400 nm. By contrast, the vd state emerges in these particles only at diameters $D$ ≥ 200 nm, where it becomes the ground state, i.e. the state with the lowest total energy. Similarly, for particles of composition TM50, the vz state competes in energy with the uniform uxy state, while the vd state appears only at larger diameters.

As shown in Figs. 3c–3e, this situation is quite typical, since in many cases the single-domain diameter $D_c$ of a particle is not determined by the state that has the lowest energy in the range $D > D_c$. For instance, Fig. 3c demonstrates that for spherical TM40 particles the single-domain diameter is determined by the vc state, which, near $D_c$ = 140 nm is stable only within a limited range of diameters: 140 ≤ $D$ ≤ 200 nm. At the same time, the vd state emerges in this particle at $D$ ≥160 nm and becomes the lowest-energy state for $D$ > 180 nm. The same situation occurs for spherical particles of composition TM50. Only in the case of spherical TM60 particles, which have positive cubic anisotropy constants, do the vz and vd vortices nearly coexist and possess very close energies.

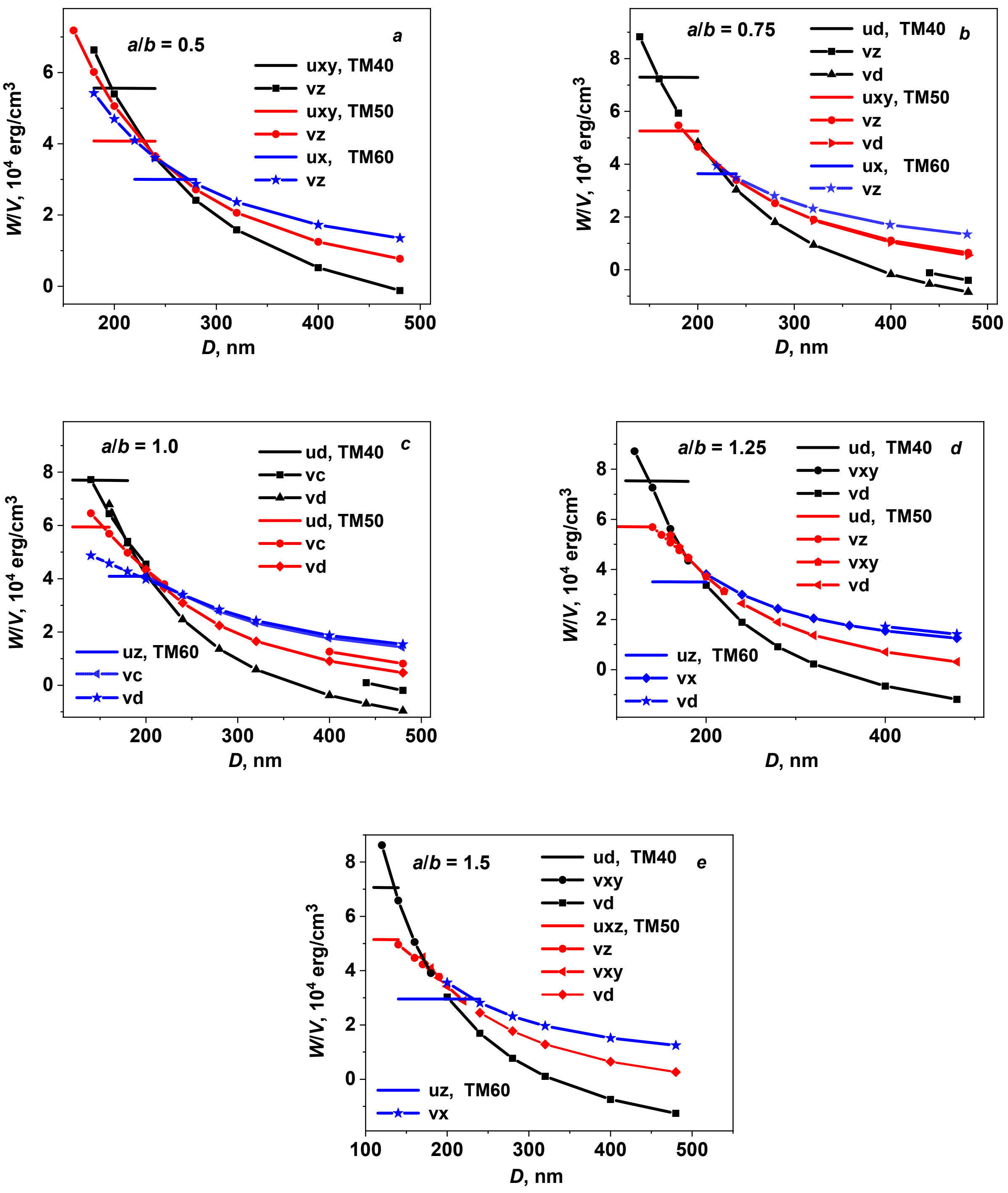

Fig. 3. Energy diagrams of stable micromagnetic states in spheroidal titanomagnetite particles with varying titanium content and different semi-axis ratios: a) $a/b$ = 0.5; b) $a/b$ = 0.75; c) $a/b$ = 1.0; d) $a/b$ = 1.25; e) $a/b$ = 1.5.

Figs. 3d and 3e show that for TM40 particles with aspect ratios $a/b$ = 1.25 and 1.5, the single-domain diameter is determined by the vxy state. The latter transforms into the vd state as the particle diameter increases. For TM60 particles with the same aspect ratios, the single-domain diameter is determined by the vx state. Meanwhile, for particles with composition TM50 and aspect ratios $a/b$ = 1.25 and 1.5, the single-domain diameter is determined by the vz state, although at larger diameters, the vxy and vd states become the lowest-energy states, respectively.

Table 2 presents the single-domain transverse diameters $D_c$(X, $a/b$) for spheroidal titanomagnetite particles, derived from an analysis of the energy diagrams shown in Fig. 3. Interestingly, the data indicate that for TM40 and TM50 compositions the single-domain diameter decreases monotonically as the particle aspect ratio increases within the range $a/b$ = 0.5–1.5, whereas for composition TM60, it first decreases and then increases. For a given aspect ratio $a/b$, the single-domain diameter increases when moving from composition TM40 to TM60. This is attributed to a significant reduction in the saturation magnetization of the particles associated with this change in composition. It should also be noted that, as is typical for magnetically soft particles, the single-domain radius of spheroidal titanomagnetite particles, defined as $R_c = D_c/2$, is of the order of the corresponding values of the exchange length $L_{ex}$.

Table 2. Single-domain diameters of TM40–TM60 particles

| | $a/b$ = 0.5 | $a/b$ = 0.75 | $a/b$ = 1.0 | $a/b$ = 1.25 | $a/b$ = 1.5 |
|---|---|---|---|---|---|
| TM40 | $D_c$ = 197 nm | $D_c$ = 160 nm | $D_c$ = 140 nm | $D_c$ = 136 nm | $D_c$ = 135 nm |
| TM50 | $D_c$ = 228 nm | $D_c$ = 185 nm | $D_c$ = 153 nm | $D_c$ = 140 nm | $D_c$ = 135 nm |
| TM60 | $D_c$ = 273 nm | $D_c$ = 232 nm | $D_c$ = 200 nm | $D_c$ = 215 nm | $D_c$ = 232 nm |

## 4. Remanent magnetization of dilute assemblies

Fig. 4 shows the magnitude of the reduced magnetic moment $M/M_s$ as a function of transverse particle diameter for the studied titanomagnetite particles having varying titanium content and aspect ratios $a/b$ = 0.5 – 1.5. The particles are in the SV states illustrated in Figs. 1 and 2. It is well known [21,30] that for diameters $D > D_c$ a creation of SV state results in a significant reduction in total energy because the curling of the magnetization vector leads to a substantial decrease in the particle's magnetostatic energy.

As shown in Fig. 4, in all the cases examined the magnetic moment of a titanomagnetite particle decreases according to an inverse power law as its diameter increases. This is a consequence of the fact that the magnetic moment of a particle in a vortex state is determined by the volume of the approximately uniformly magnetized vortex core, because in the outer region of the particle the unit magnetization vector possesses a significant circular component which contributes nothing to the particle's total magnetic moment. Meanwhile, the radius of the vortex core changes only slowly as the particle diameter increases [30].

According to Fig. 4, different types of SV states with various magnetic moments can coexist in certain cases within the same particle. For a given geometry, the total magnetic moment of a titanomagnetite particle is at its minimum and at its maximum for the TM40 and TM60 compositions, respectively. This is primarily due to the significant difference in the saturation magnetization of these compositions. A higher saturation magnetization $M_s$ for the TM40 composition leads to a minimization of the particle's total energy at a smaller vortex core radius. This, in turn, results in a reduction of the particle's magnetic moment, which is proportional to the square of the vortex core radius.

It is worth noting that for particles with a sufficiently large diameter the reduced particle magnetic moment reaches very low values. For instance, for TM40 particle with a diameter $D$ = 480 nm the reduced magnetic moment is $M/M_s$ = 0.0083 at aspect ratio $a/b$ = 0.5, however it reaches the value $M/M_s$ = 0.0429 at $a/b$ = 1.5. Thus, for a fixed diameter, the magnetic moment of a particle increases with an increase in its aspect ratio. This is associated with an increase in the length of the vortex core.

As noted above, for the vz state the magnetic moment vector can take only two orientations: parallel or antiparallel to the particle's symmetry axis $Z$. This means that for the vz state the corresponding curves in Fig. 4 provide the remanent magnetization of dilute, oriented assemblies in cases when the external magnetic field is parallel to the assembly's orientation axis. To estimate the remanent magnetization of an assembly characterized by a random, spatially uniform distribution of particle symmetry axes it is necessary to take into account that when external magnetic field is reduced from a large positive value to zero, the projection of the particle magnetic moment on the applied field direction remains positive. As a result, for randomly oriented assemblies the magnetic moment values for the vz state in Fig. 4 must be multiplied by a factor $\eta = 0.5$. This is because, for randomly oriented assembly the non-equivalent directions of the external magnetic field with respect to particle symmetry axis span the upper hemisphere with the solid angle $2\pi$. An ambiguity arises only when, for a given diameter, different SV states can coexist within a single particle. As illustrated in Figs. 4b, 4d, and 4e, such a scenario can occur, for instance, for vz and vd states in TM50 particles within certain diameter ranges.

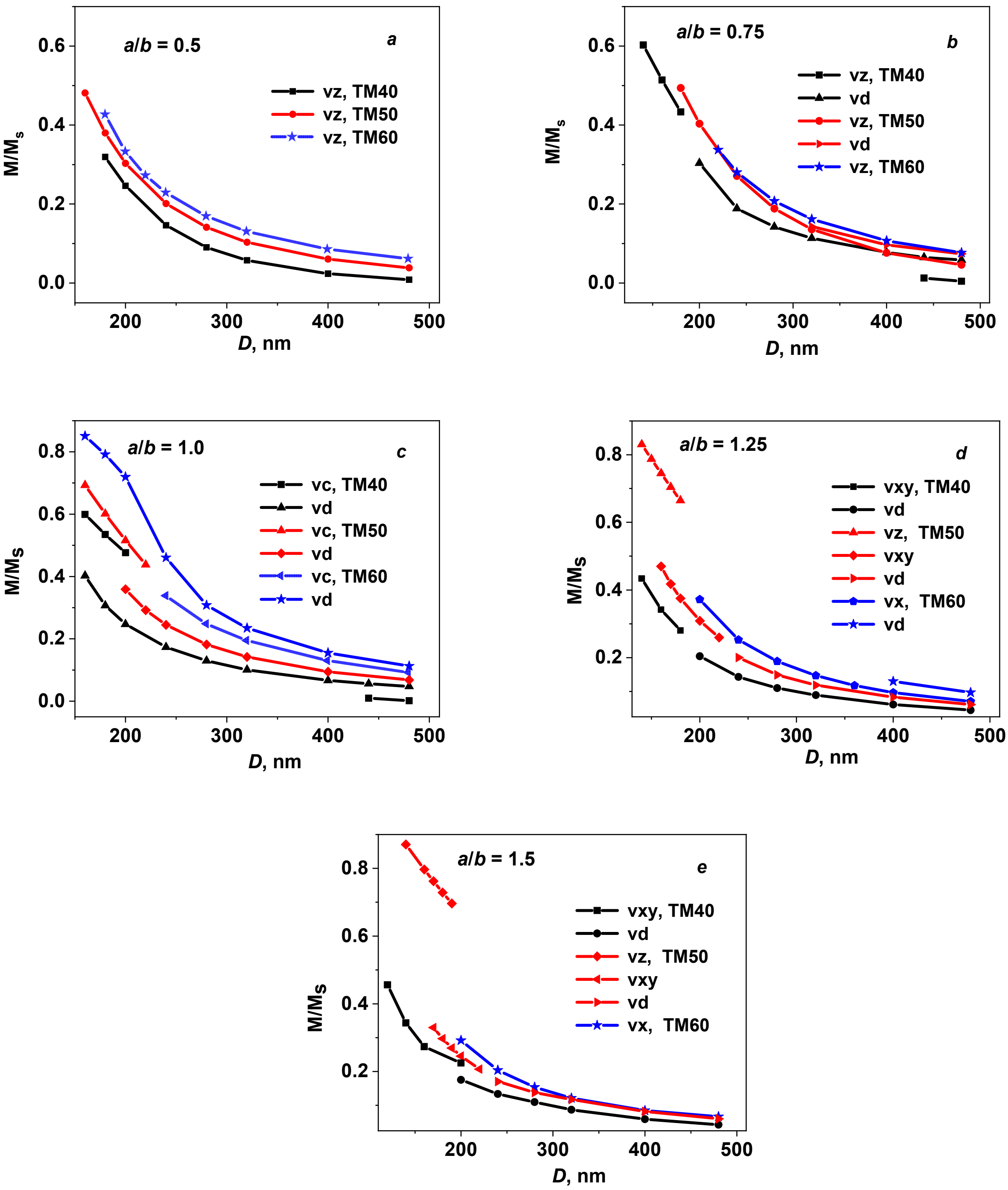

Fig. 4. Reduced magnetic moments of titanomagnetite particles in different SV states for various aspect ratios: a) $a/b = 0.5$; b) $a/b = 0.75$; c) $a/b = 1.0$; d) $a/b = 1.25$; e) $a/b = 1.5$.

It is well known [29, 31] that for spherical particles with cubic anisotropy, the corresponding conversion factors for determining the remanent magnetization of a randomly oriented assembly are $\eta$ = 0.831 for vc states and $\eta$ = 0.866 for vd states, respectively.

For the vx state in TM60 particles there are four equivalent orientations of the particle's magnetic moment, all parallel or anti-parallel to the Cartesian *X* and *Y* axes. Consequently, the region of irreducible orientations Ω for an external magnetic field spans a solid angle π and can be defined in spherical coordinates ($\theta$, $\varphi$) by the inequalities $0 \le \theta \le \pi$, $|\varphi| \le \pi/4$. By calculating the projection of unit vector along the external magnetic field onto the *X*-axis, averaged over the domain Ω, one obtains for this case a conversion factor $\eta = \sqrt{2}/2 \approx 0.707$. The same conversion factor applies to the vxy state, which differs from the vx state only in that the particle's magnetic moment has projections onto the *X* and *Y* axes with equal absolute values, $|m_x| = |m_y|$.

Finally, due to the symmetry of equation (3), for vd states there are 8 equivalent configurations with identical positive and negative projections of the moment onto the *Z*-axis and equal absolute values of the projections onto the *X* and *Y* axes, $|m_x| = |m_y|$. Therefore, the domain Ω of irreducible directions for the external magnetic field spans a solid angle π/2 and can be defined in spherical coordinates by the inequalities $0 \le \theta \le \pi/2$, $0 \le \varphi \le \pi/2$. Let the equilibrium moment of a particle in the vd state, in the absence of external magnetic field, be parallel to a direction $\boldsymbol{n} = \left(\sin\theta_0/\sqrt{2}, \sin\theta_0/\sqrt{2}, \cos\theta_0\right)$. By calculating the average of the projection of unit vector along the external magnetic field onto the vector $\boldsymbol{n}$ one obtains for the vd state a conversion factor $\eta = \sin\theta_0/\sqrt{2} + \cos\theta_0/2$. In the specific case of vd state in spherical particle, setting $\cos\theta_0 = 1/\sqrt{3}$, one obtains from this equation the well-known value $\eta = 1.5/\sqrt{3} \approx 0.866$.

Calculations show that the magnetic dipole interaction between submicron-sized titanomagnetite particles being in SV states is extremely weak, even when the particles are in close proximity, at distances on the order of their radius. This is due to the small magnitude of the particles' total magnetic moments. Consequently, shown in Fig. 4 results, obtained for dilute assemblies of spheroidal titanomagnetite particles, should also hold for assemblies with a significantly high particle density.

It was experimentally demonstrated [20] that, in prismatic submicron titanomagnetite particles, SV states exhibit high thermal stability, a property essential for the preservation of a stable paleomagnetic remanent magnetization. However, as shown above, vortex states in spheroidal submicron titanomagnetite particles possess several equivalent orientations for the vortex core, separated by energy barriers. Consequently, transitions between these equivalent vortex states can occur under the influence of thermal or magnetic perturbations. Therefore, the stability of the remanent magnetization in an assembly of spheroidal titanomagnetite particles is worth further investigation.

## 5. Conclusions

Calculations performed reveal a wide variety of stable SV states in spheroidal titanomagnetite particles with aspect ratio $a/b$ = 0.5–1.5 and transverse diameters ranging from 150 to 480 nm. The study focuses on particles with compositions in the TM40–TM60 range, which are most commonly found in ancient volcanic rocks. For each particle type, the transverse diameters defining the single-domain threshold were determined by comparing the total energies of uniform and vortex states. It is shown that the remanent magnetization of titanomagnetite particle assemblies decreases according to inverse power law as the transverse particle diameter increases. For titanomagnetite particles of highest diameters studied, $D \sim 500$ nm, the normalized remanent magnetization of an assembly can reach very low values, $M/M_s < 0.01$.

It should be noted that in this work the calculations of magnetization distributions were performed neglecting for the magnetostriction energy, which can be significant in titanomagnetite particles of the compositions studied [32, 33]. To our opinion, incorporating the effects of magnetostriction is a separate task, as it would require accounting for mechanical quenching stresses present in natural volcanic rock samples. The presence of SV states should significantly influence the properties of titanomagnetite particle assemblies characterized by broad distributions in composition, size, and shape, as it is typical for the dispersed natural assemblies found in volcanic rocks [4–6]. The results obtained in this work should facilitate a more accurate interpretation of contemporary paleomagnetic data.

**Conflict of interest**

The authors declare that they have no conflict of interest.